\documentclass[%
 reprint,
 amsmath,amssymb,
 aps,
 showkeys,
]{revtex4-1}

\usepackage{graphicx}
\usepackage{dcolumn}
\usepackage{bm}
\usepackage{graphicx}
\usepackage{tikz}
\usetikzlibrary{shapes}
\usepackage{pgfplots}
\usepackage{float}
\usetikzlibrary{shapes,snakes,arrows.meta, matrix, positioning}
\usetikzlibrary{arrows,decorations,backgrounds,positioning,fit,petri}
\usepackage{bbm}
\usepackage{scalerel,stackengine}
\stackMath
\newcommand\reallywidehat[1]{%
	\savestack{\tmpbox}{\stretchto{%
			\scaleto{%
				\scalerel*[\widthof{\ensuremath{#1}}]{\kern-.8pt\bigwedge\kern-.8pt}%
				{\rule[-\textheight/2]{1ex}{\textheight}}
			}{\textheight}%
		}{0.8ex}}%
	\stackon[1pt]{#1}{\tmpbox}%
}

\begin{document}


\title{Bayesian parameter estimation of miss-specified models}

\author{Johannes Oberpriller}
\author{Torsten En\ss lin}%
\affiliation{%
 Max-Planck-Institut f\"ur Astrophysik, Karl-Schwarzschildstr. 1, 85748 Garching, Germany \\
 Ludwig-Maximilians-Universit\"at M\"unchen, Geschwister-Scholl-Platz 1, 80539 Munich, Germany
}%

\begin{abstract}
Fitting a simplifying model with several parameters to real data of complex objects is a highly nontrivial task, but enables the possibility to get insights into the objects physics.
Here, we present a method to infer the parameters of the model, the model error as well as the statistics of the model error. 
This method relies on the usage of many data sets in a simultaneous analysis in order to overcome the problems caused by the degeneracy between model parameters and model error. 
Errors in the modeling of the measurement instrument can be absorbed in the model error allowing for applications with complex instruments.
\end{abstract}

\pacs{Valid PACS appear here}
\maketitle


\section{Introduction}\label{intro}
\subsection{General problem}
%
Whenever phenomena of nature are not directly accessible, it is a common way to build a model of them. 
These models usually have parameters for which one has to find the best fitting parameter values for each phenomena.
Unfortunately, in this process of fitting and modeling there are two main problems.
On the one hand, models are often derived from theoretical analysis of the underlying processes. 
However, these theoretical models describing complex real world systems are necessarily idealizations.
The implied simplifications allow us to focus on the essential physics, to keep the model computationally feasible, and to investigate systems for which not all of their components are perfectly known. 
On the other hand there is always measurement noise in the observation process.

By judging theoretical models on a goodness-of-fit basis they often perform insufficiently when faced to real data of the actual system.
The difference between real and modeled data can easily exceed the error budget of the measurement.
The reason is that the idealized models do not capture all aspects of the real systems. 
But aspects, which are not modeled, imprint on the data as well.
To discriminate these from measurement errors or noise, we use the term $model \, errors$ to describe these imperfections of our theoretical description.
A common approach to determine the model parameters from the data is a likelihood based methodology ($\mathcal{X}^{2}$-fit \cite{leastsquare, legendre2018nouvelles}, maximum likelihood \cite{MLmisp_mod}, or Bayesian parameter estimation \cite{bay_para_est}) that uses the measurement uncertainties as a metric in data space. 
The resulting parameter estimates can be strongly distorted by the desire of the method to minimize all apparent differences between predicted and real data, indifferently if these differences occur due to measurement or model errors. 

Thus, the model error should be included into the error budget of the parameter estimation as well.
However, this requires a model for the not yet captured aspects of the system or at least for the model errors which are produced by the missing aspects.
In a rigorous way, we need to do a case by case analysis of the missing physics. 
However, this would be completely impractical in cases like the one we have in mind - spectroscopic investigation of complex astrophysical objects.
Instead, this paper aims to construct a plausible, but by no means perfect, effective description of the model errors and their uncertainty. 
We want to show how the estimation of the model parameters is improved by taking a model error into account. 
\subsection{Example of the general problem in astrophysics: Spectroscopic investigation of complex astrophysical objects}
``\textit{Remember that all models are wrong, the practical question is how wrong do they have to be to not be useful.}"\cite{box1987empirical}.
This quote of George E.P. Box, a frontier in Bayesian inference, is well applicable for astrophysical models.
An example for the process of modeling and fitting is the task to fit stellar models to observed stellar spectra.
Many different physical conditions determine the spectrum of a star e.g. stellar parameters like the effective temperature $T_{\mathrm{eff}}$, the magnitude of the gravitational surface acceleration $\mathrm{log}\,g$, the metalicity $[\mathrm{Fe/H}]$ and abundances for almost the full periodic table.  
A usual approach to determine stellar conditions is fitting data to synthetic model spectra. \\
The calculations of synthetic spectral models are normally done by radiative transfer calculations.
However, if one accounts for all physical conditions in these calculations one can not keep the models computationally feasible.
Thus, some simplifications have to be done e.g. local thermodynamical equilibrium (LTE) or plane parallel geometry.
In the resulting star models only the most important physical parameters are included.\\
We will denote physical conditions included in the process of modeling with $p$, while not modeled physical conditions will be denoted with $c$.
Additionally, it is still very hard to produce high quality synthetic data due to inaccuracies in atomic and molecular constants.
Thus, it is almost impossible to ensure spectral lines to fit perfectly over all stellar parameters and spectral ranges. \\
Due to the inaccuracy in the modeling process stellar spectra can only be described approximately. 
As an illustrative, simple model for an absorption spectrum accounting for a gray body spectrum with absorption lines, one could use the following: 
\begin{equation}\label{eq: model2}
\begin{aligned}
t_{\nu}(p)&= A \left( \frac{\nu}{\nu_{0}}  \right)^{\alpha} e^{-\frac{\nu}{T_{\mathrm{eff}}}} e^{-\tau_{\nu}} \, , \, \mathrm{with} \\
\tau_{\nu} &= \sum_{i = 1}^{3} t_{i} \, \mathcal{G}(\nu-\nu_{i},\sigma_{i}^{2}) \,  ,
\end{aligned}
\end{equation}
where $\mathcal{G}(\nu-\nu_{i},\sigma_{i}^{2})$ denotes a Gaussian in $\nu$ with mean $\nu_{i}$ and standard deviation $\sigma_{i}$.
The exponent of the power-law $\alpha$, the effective temperature $T_{\mathrm{eff}}$ and the strength parameters for the absorption line spectrum $t_{1}, t_{2},t_{3}$ are assumed to be the unknown parameters, while the amplitude $A$, the mean frequency $\nu_{0}$, the means of the absorption peaks $\nu_{1},\nu_{2},\nu_{3}$, and the line widths $\sigma_{1}^{2},\sigma_{2}^{2},\sigma_{3}^{2}$ are known.
This simplyfing model does not account for the gravitational surface acceleration $\mathrm{log}\,g$, the metalicity $[\mathrm{Fe/H}]$ and all abundances of the periodic table expect for three. 

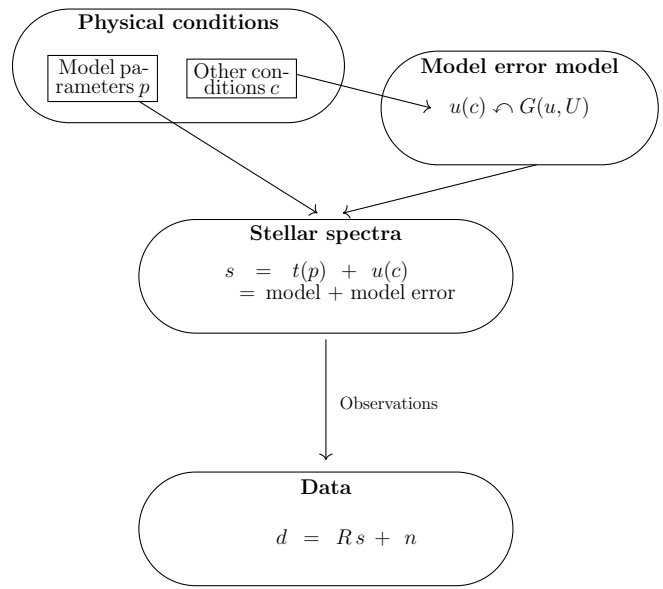
\begin{figure}[]
	\def\myellipse{(-3.5,0) ellipse (3.5cm and 1.7cm)}
	\def\mysecondellipse{(0,-5) ellipse (3cm and 1.5cm)}
	\def\mythirdellipse{(0,-11) ellipse (3cm and 1.5cm)}
	\def\myfourthellipse{(4.6,-1) ellipse (3.5cm and 1.7cm)}
	\begin{tikzpicture}[scale = 0.56, every node/.style ={scale = 0.56}]
	\node[place, align = center, inner sep = 0.1cm] at (4.6,-1) (myfourthellipse) [shape = rounded rectangle, text width = 4.7cm, text height = 2.5 cm]{};
	\node[place, align = center] at (4.6, 0) (model error model)[draw = none]{\Large \textbf{Model error model}};
	\node[place] at (4.6, -1) (uU)[draw = none]{\Large $u(c) \curvearrowleft G(u,U)$};
	\node[place, align = center, inner sep = 0.1cm] at (-3.5,0) (myellipse) [shape = rounded rectangle, text width = 6cm, text height = 2.5 cm]{};
	\node[place,align = center, inner sep = 0.1cm] at (-5.3,-0.3)  (modelparameters)[shape= rectangle, text width = 2.4cm]  {\,\Large Model parameters $p$};
	\node[place, align = center, inner sep = 0.1cm] at (-3.5, 1) (physicalconditions)[draw = none]{\Large \textbf{Physical conditions}};
	\node[place, align = center, inner sep = 0.1cm] at (-2.,-0.3)  (otherconditions)[shape=rectangle, text width = 2.4cm]  {\Large \centering Other conditions $c$};
	\node[place, align = center, inner sep = 0.1cm] at (0,-5) (mysecondellipse) [shape = rounded rectangle, text width = 7cm, text height = 2.5 cm]{};
	\node[place] at (0.6, -5.1) (stellarspectramath)[draw = none, text width = 6cm]{\Large{$ s  =  t(p) \,  + \,  u (c)$ \\  $ \, \, \, \,  = \mathrm{model} + \mathrm{model\;error}$}};
	\node[place] at (0, -4) (stellarspectra) [draw = none] {\Large \textbf{Stellar spectra}};
	\node[place, align = center, inner sep = 0.1cm] at (0,-11) (mythirdellipse) [shape = rounded rectangle, text width = 7cm, text height = 2.5 cm]{};
	\node[place] at (0.4, -11.2) (datamath)[draw = none, text width = 4cm]{\Large $ \, \, \, \, \, d  \; =  \; R \,s \,  +   \;n$ };
	\node[place] at (0, -10) (data) [draw = none] {\Large \textbf{Data}};
	\node[place]  (placeholder)[shape= coordinate, above = 0.002cm of stellarspectra]{};
	\draw[->]	(modelparameters) to (-0.2,-3.5);
	\draw[->]	(0,-6.5) to (data);
	\draw[->]   (-0.7,-0.2) to (2.5,-1);
	\draw[->]   (5,-2.35) to (0.4,-3.5);
	\node[place] at (1.5, -8) (data) [draw = none] {\large Observations};
	\end{tikzpicture}
	\caption{Systematic representation of the connection between physical conditions on the star and an model error due to not modeled conditions.}\label{systematic_model}
\end{figure}

Assuming one determines the correct values $p$ of the star models, not modeled conditions will present themselves in a model error.
However, we have observations of many stars.
If some of them are fitted with the same model, this gives us a chance to at least identify and characterize from star to star varying spectral structures due to model errors.  
Then, the model errors can be statistically added to the error budget.
However, parameter determination and determination of the statistics of the model errors are coupled, which is the reason why they must be performed together.
Here we use a generic, hierarchical Bayesian model based on Gaussian processes to describe the model errors, which is schematically shown in fig.~\ref{systematic_model}.
\\
The structure of this work is the following.
In sec.~\ref{state} we explain the general problem of parameter estimations of miss-specified models and give an overview of the state of the art methods tackling this problem. 
A brief discussion of the Wiener filter given in sec.~\ref{Wiener_filter}, since the understanding of it is crucial for our algorithms. 
In sec.~\ref{model_error_model} we state and explain our model for the model error and its uncertainty and furthermore discuss our prior assumptions.
In sec.~\ref{method} we derive and present the inference algorithm and show its capability on mock data.
Finally, we compare the inferred results with a least-square parameter estimation. 
\section{Problem setting and state of the art}\label{state}
Computer models simulating star spectra are invaluable tools in astrophysics.
Unfortunately, there is typically uncertainty in the predictions of the synthetic star spectra models.
The arising errors can be categorized in two main sources.
The first one results from the use of an inadequate model with a correct set of parameters values.
This means one has found the true parameter values for the wrong model, which is called the model error. 
The second one is the uncertainty in the parameters considering the true model, which is called the \textit{parameter error}. 
By modeling and computing stellar spectra both errors are present simultaneously. 

Let us denote with $t^{\ast}$ the true forward operator of the model $\mathcal{M}$ and with $t$ another forward operator, e.g. the forward operator used for the parameter estimations.
Let $p^{\ast}$ be the true set of parameter values and $p$ another set of parameter values. 
Then the parameter error is the difference between $t^{\ast}(p)$ and $t^{\ast}(p^{\ast})$
\begin{equation}
u_{\mathrm{par}} = t^{\ast}(p^{\ast}) - t^{\ast}(p)\, ,
\end{equation}
which is independent of $t$, but a quantification would require the knowledge of $t^{\ast}$. 
However, $t^{\ast}$ is not available to us as it represents the true physics.
This raises the question of a meaningful parameter error evaluation, especially in the absence of a model error model.

The direct evaluation of model errors, defined as 
\begin{equation}
u_{\mathrm{model}} = t^{\ast}(p^{\ast}) - t(p^{\ast}) \, ,
\end{equation} is a subject of large uncertainty and has only little use in practice.
The importance of model errors in parameter estimation is widely recognized, but there exists only little research about it. 
Thus, still no general methodology for quantifying the impact of model error for parameter estimations and its influence on parameter error analysis is available.
 
In literature a common way to describe the forward problem used to estimate the unknown parameters is described in the following way
\begin{equation}\label{eq: forward_problem}
d = t(p) + e \, , 
\end{equation}
where $d$ is the observed data, $t(p)$ the forward operator belonging to the mathematical model $\mathcal{M}$ and $e$ is the error. 
The error $e$ is the sum of a measurement error $n$ and a model error $u$
\begin{equation}
e = n + u \, .
\end{equation}
The inverse problem is formulated by finding the best fitting parameters $p$.
Usually this is achieved by minimizing the residual $e$ via a classical or weighted least-square method.
In this method $p$ is obtained by minimizing the sum of the squared differences
\begin{equation}\label{eq : least_square}
\hat{p} = \min \limits_{p} \, \left[d-t(p)\right]^{\dagger}E^{-1}\left[ d-t(p)\right] \, ,
\end{equation}
where $E$ represents the covariance of the error $e$. 
The underlying assumption for this formulation of least-square is that the measurement error and model error have the same statistics. 
However, there are two main problems.

On the one hand the model error is not random, but it is systematic and certainly will not have the same statistic as the measurement error.
The model error will for different measurements of objects, which are described by the same model and same parameters, have a comparable strength on different positions and a certain correlation structure. 
On the other hand the covariance of the model error is not known a priori and can only be estimated. 
Thus, this approach is almost useless, if the model error $u$ is dominating the overall error $e$.

Another main problem in estimating parameters and model error is that several approaches do not give a unique solution for the parameters $p$ due to the degeneracy between the model error and the parameters.
One way to tackle the problem of degeneracy is adding prior information on the parameters.
The objective function minimized to find the best fitting parameters reads
\begin{equation}\label{eq: maping}
\mathcal{H} = \left[d-t(p)\right]^{\dagger}E^{-1}\left[ d-t(p)\right] + (p-p_{e})^{\dagger}P^{-1}(p-p_{e}) \, ,
\end{equation}
where $P$ is the covariance matrix for the parameters and $p_{e}$ is a estimated set of parameter values. 
The assumption for the parameters is that they are Gaussian distributed. 
This approach allows to derive a posterior uncertainty of the parameters $p$, where all plausible sets of parameter values should be represented.
Unfortunately, the model error is again not incorporated directly in the modeling process for the data observation and this approach has the same drawbacks as least-square. 

Assuming we know the measurement error distribution, we can write the solutions for the inverse problem for $d= t(p) +u +n$ as a set
\begin{equation}\label{eq : manifold}
M_{s} = \{ (p, u): s = t(p) + u  \} \, ,
\end{equation}
where $s$ is the quantity of interest.
This shows that the degeneracy of the model error and the parameters can be embedded on a manifold, on which the solutions are equally distributed without prior knowledge.

The authors of \cite{2014Hagan} investigated the role of the priors on the model parameters and model errors in the inference.
For one data set, all points of $M_{s}$ have equal likelihood and the posterior distribution on this manifold is completely determined by the prior.
The probability to obtain the true values of $p$ and respectively $u$ will be high, if the prior has a high density for the correct solutions relative to the other solutions. 
In general one can distinguish two extreme cases.

For the first case, the authors of this study considered a weak prior on $p$ and a diffuse, zero-mean prior distribution for $u$.
It was found by \cite{2001Hagan}, that the posterior will be flat over the set of $(p,u)$ that result in the same $s$.
As the authors of ~\cite{2001Hagan} assumed a zero-mean prior on $u$, the probability in regions, where $u$ is near zero, will be higher.
Thus, the estimated model parameters $p$ will be similar to a maximum a posteriori estimator, which may not cover the true values of $p$.

The second case is given, if realistic prior information for $u$ and/or $p$ is available, which leads to a tighter posterior distribution. 
Unfortunately, if the prior information is wrong, a low probability is attached to the true value of $p$ and the posterior uncertainty may fail to cover them.
If a strong prior for the parameters is available from other observations that do not rely on the model process, the quality of model error and model parameter estimation is higher.

As we have seen, there are many subtleties in these methods.
Thus, we will now discuss the main problems of approaches in literature to tackle the issue of a miss-specified model and give a conceptual overview of our ideas.
If we try to isolate the effect of the model errors for the parameter estimation with the help of eq.~\ref{eq: maping} this will lead to following problems. 

The first problem is that the model error is not random and does certainly not have the same probabilistic statistics as the measurement error. 
To solve this issue we will try to model the model error as its own quantity with its own statistic.

The second problem is that although the inclusion of prior information on the parameters in the objective function reduces the ill-posedness of the inverse problem, it does not guarantee the uniqueness of the solution. 
As we are in the comfortable situation to have more than one data set with approximately the same parameter values for our analysis, the ill-posedness can be reduced further.

The third problem is that each data set can be fitted in different ways. 
Thus, one parameter set can be the best fit for the first data set and another set of parameter values can be the most appropriate for the next data set. 
We will analyze all available data sets at the same time, but do not assume that all data sets have the same true parameters. 
Note that this is not the same as combining the results of an analysis of each data set on its own, since the different data sets share the same model error statistics. \\ 
\section{Wiener filter theory}\label{Wiener_filter}

In the following, we want to give a short overview of the Wiener filter.
Stellar spectra, which we will also call the signal $s$ in this work, are physical fields.
Thus they have infinitely many degrees of freedom, which motivates the usage of the framework of information field theory (IFT)~\cite{IFT} in which the Wiener filter appears as the solution to the free field case.
We observe a single stellar spectrum with the help of a measurement instrument that somehow discretizes a field, e.g. by integrating and averaging. 
In our mathematical description we represent the measurement instrument as a linear operator $R$.
Further, we assume the instrument to be well known, such that errors arising from the modeling process of the measurement instrument will be ignored.
%
%
%
The measured data $d$ can be expressed by applying $R$ to the signal $s$ corrupted with noise $n$.
The measurement equation describing this reads
\begin{equation} \label{data-equation}
d = R\,s+n \, .
\end{equation}
In particular, we assume the noise $n$ to be Gaussian, i.e.
\begin{align}
	n \,\, &{\curvearrowleft} \,\, \mathcal{G}(n,N) \\
	\mathcal{G}(n,N) &= \frac{1}{\vert 2 \pi N \vert }e^{-\frac{1}{2}n^{ \dagger} N^{-1}n} \, ,
\end{align}
where we denote with $n^{\dagger}$ the complex conjugate and adjoint of $n$ and $N$ is representing the covariance of the noise distribution.
Mathematical conventions in this formula, e.g. how an operator acts on a field and how the scalar product is defined, are explained in appendix~\ref{definitions}. \\
By using the measurement eq.~\ref{data-equation} the likelihood for the data given the signal and noise realization can be derived.
The conditional probability $\mathcal{P}(d\vert s,n)= \delta[d-(Rs+n)]$ tells us how probably the data is given the noise and the signal:
\begin{align}
\mathcal{P}(d\vert s,n)= \delta[d-(Rs+n)] \, .
\end{align}
Since we do not know the noise, we marginalize it out:
\begin{equation}
\begin{aligned}
\mathcal{P}(d\vert s) &= \int \mathcal{D}\, n \, \mathcal{P}(d,n\vert s) = \int \mathcal{D}\,n \,\mathcal{P}(d\vert s,n)\, \mathcal{P}(n) \\
&=\mathcal{G}(d-R\,s,N) \, .
\end{aligned}
\end{equation}
For the task of infering the signal $s$ from the data $d$, we have to inspect the posterior distribution. 
This is provided by Bayes theorem \cite{Bayes_rule},
\begin{equation}
\mathcal{P}(s|d) = \frac{\mathcal{P}(d \vert s)\mathcal{P}(s)}{\mathcal{P}(d)} \, .
\end{equation}
In analogy to statistical physics, Bayes theorem is reformulated by introducing the information Hamiltonian $\mathcal{H}(s, d)$ and the partition function $\mathcal{Z}(d)$
\begin{align}
\mathcal{P}(s|d) &= \frac{\mathcal{P}(d \vert s)\mathcal{P}(s)}{\mathcal{P}(d)}= \frac{1}{\mathcal{Z}(d)}e^{-\mathcal{H}(d,s)},\, \, \mbox{with} \label{statistical_physics} \\
\mathcal{H}(s, d) &= -\ln[P(s,d)] \, , \\
\mathcal{Z}(d)&= \int \mathrm{d}s \, e^{-\mathcal{H}(s,d)} = \mathcal{P}(d) .
\end{align}
We will assume a Gaussian prior for the signal $s$, i.e.
\begin{equation}
\mathcal{P}(s) = \mathcal{G}(s,S) ,
\end{equation}
where for the moment the prior covariance $S$ of the signal $s$ is assumed to be known.
Together with the data model and Bayes equation, this allows us to calculate the information Hamiltonian $\mathcal{H}(s,d)$, which is given by 
\begin{equation}
\begin{split}
\mathcal{H}(s,d) &= \mathcal{H}(d \vert s) + \mathcal{H}(s) = \frac{1}{2}(d-Rs)^\dagger N^{-1}(d-Rs) \\
&+ \frac{1}{2}s^\dagger S^{-1}s + 
\frac{1}{2} \ln \vert 2\pi N \vert + \frac{1}{2} \ln \vert 2\pi S \vert \, .
\end{split}
\end{equation}
A quadratic completion leads to a more conventional form of the information Hamiltonian, where we drop terms not depending on $s$, as denoted by the equality up to constants "$\widehat{ =}$":
\begin{align}
\mathcal{H}(s,d) \, &\widehat{ =} \, \frac{1}{2}\, (s-m)^{\dagger} \, D^{-1}\, (s-m)\, , \\
D &= (S^{-1} + R^\dagger N^{-1} R)^{-1}  \label{Wiener_filter_curvature} \, ,\\ 
j &= R^\dagger N^{-1}d\, , \\
m &= Dj \, .	
\end{align}
As mentioned before, the posterior is the quantity of interest in the inference and with a short calculation we get 
\begin{equation}\label{proportionalities}
\mathcal{P}(s \vert d) \propto e^{-H(s,d)} \propto e^{-\frac{1}{2}(s-m)D^{-1}(s-m)} \propto \mathcal{G}(s-m,D) \, .
\end{equation}
Since the l.h.s and the r.h.s. of eq.~\ref{proportionalities} are normalized,
\begin{equation} 
\mathcal{P}(s \vert d) = \mathcal{G}(s-m, D)
\end{equation}
holds exactly. 
Here we introduced the mean $m = DR^{\dagger}N^{-1} d$, which can be written as applying the information propagator $D = \left( S^{-1} + R^{\dagger}N^{-1}R \right)^{-1}$ on the information source $j = R^{\dagger}N^{-1}d$. 
The operation $DR^{\dagger}N^{-1}$  is known as the Wiener filter \cite{Wiener}. \\
In many real world cases the prior uncertainty structure of the signal $s$ and additional fields are not known a priori and one would like to obtain posterior uncertainties.
One way to counteract these problems is proposed in~\cite{2017arXiv171102955K,Global-Newton} which reconstructs the uncertainty structure alongside the signal simultaneously. 
We will adopt this improved formulation in the following.
\section{Model error model}\label{model_error_model}
In this section we motivate our model for the model error and its uncertainty.
Furthermore, we want to derive the information Hamiltonian for this specific choice of model error uncertainty. 
As our goal is to identify the model parameters of a theoretical model, we have to build a model for the model error.
Additionally, we have to connect the model error to observed data. \\
Assuming we have spectral observations of $n$ stars, we denote the spectral data of star $i$ with $d_{i}$ and the data vector $d= (d_{i}, \cdots , d_{n} ) $ describes the ordered collection of all data sets in the data vector. 
The measurement equation is given in eq.~\ref{data-equation},
where the signal can be described as a parameter model $t(p)$ corrupted with a model error $u$.
However, in the case of a spectroscopic analysis of an absorption spectrum, we want to ensure, that the signal is positive. 
Therefore, we adopt the following measurement model:
\begin{align}
s &= t(p)+u \, ,\\
d &= R \, e^{\left[t(p) + u \right]}  + n\, . \label{real_data_equation}
\end{align}
This change to a positive definite signal has two main consequences. 
On the one hand the model error $e^{u}$ is now multiplicative to the model $e^{t(p)}$, which is often a more accurate description than an additive model error as it allows a better separation of systematic model error and measurement error~\cite{multiplicativerror}.
On the other hand, to compare the model to data, we have to take the logarithm of the model. 
In case of our spectroscopic model in eq. \ref{eq: model2}, this is given by 
\begin{equation}\label{eq: reformulated model}
\begin{aligned}
	t(p) \rightarrow \ln[t_{\nu}(p)] &= \ln(A) + \alpha \ln( \frac{\nu}{\nu_{0}}) -\frac{\nu}{T_{\mathrm{eff}}}  \\ 
	&- \sum_{i = 1}^{3} t_{i} \, \mathcal{G}(\nu-\nu_{i},\sigma_{i}^{2}) \, .
\end{aligned}
\end{equation}
The consequence is that now the model is linear in every parameter except for $T_{\mathrm{eff}}$.

For a rigorous statistical analysis we need a prior $\mathcal{P}(p)$, which is also necessary to reduce the degeneracy between the parameters and the model error.

However, we also have to build a model for the statistics of the model errors $u$.
We assume the model error to be Gaussian with a model error uncertainty covariance $U$.
By analyzing data sets with the same model forward operator $t$ and the same model-generating process, we can assume that the model errors $u_{i}$ follow the same statistics. 
For the model error uncertainty covariance $U$ we try to be as general as possible for the reasons explained in sec.~\ref{state}.
On the one hand, we have to take care of locally varying errors, e.g. peaks in stellar spectra, which are not modeled. 
This task can be tackled with a diagonal matrix $\widehat{\,e^{\theta}\,}$ in position space (here $\nu$, parameterized by the field $\theta$). 
On the other hand we think that for some part of the model error budget a priori no position should be singled out.
Thus, statistical homogenuity is assumed, which is according to the Wiener-Khintchin theorem \cite{Wiener, Khintchin}, represented as diagonal matrix $\widehat {\,e^{{\tau}}\,}$ in Fourier space (here the Fourier domain conjugate to $\nu$).
For signals with more than one dimensions statistical isotropy might be assumed as well. 
This means to assume all directions to be statistically equal, resulting only in a dependence of that part of the model error correlation on the relative distance of the compared locations.
Thus, statistical isotropy leads to a collapse of the Fourier-diagonal part of the correlation structure to a one-dimensional power spectrum function $e^{\tau} = e^{\tau(\kappa)}$ of the harmonic modes $\kappa$. 
This is incorporated in the uncertainty through the power projection operator $\mathbb{P}$, which is explained in appendix \ref{power_proj}, and the Fourier transform operator $\mathbb{F}$.
One simple way to combine both parts of uncertainty structures into the uncertainty covariance is a modified product of both, which is given by
\begin{align}
u \,\, {\curvearrowleft}\,\, G(u,U) \, \\
U=\widehat {e^{{\theta_{x}}}}\mathbb{F}^{\dagger}\widehat{ \mathbb{P}^{\dagger} e^{{\tau_{y}}}}\mathbb{F}\widehat {e^{{\theta_{x}}}} \label{U_para} \, .
\end{align}
The index $y=\ln(k)$ indicates that we assumed a logarithmic scale for the Fourier space coordinate of $\tau$ and the index $x$ denotes a linear space coordinate for $\theta$.
We indicate the transformation of a field to a diagonal operator in the domain of the field by "$\widehat{\;\;\; \; \;}$".
In this parameterization the model error uncertainty is a multiplicative combination of a mask in position space $\widehat{e^{\theta_{x}}}$ and a convolution. 
Any field processed by $U$ (and also by its inverse) first goes through the (inverted) mask, then gets convolved (deconvolved) and at last the (inverted) mask again imprints on the field. 

Again for the sake of a statistical rigorous analysis, we need a prior for $\tau$ and $\theta$.
To account for the smoothness of the power spectrum on a logarithmic scale, a smoothness prior is adopted to $\tau$.
Similarly, a smoothness prior for $\theta$ on a linear scale in position space is adapted to $\theta$:
\begin{align}
H(\tau)=\frac{1}{2\sigma_{\tau}^{2}} \int \mathrm{d}y \left( \frac{\partial^{2}\tau}{\partial^{2}y}\right)^{2} = \frac{1}{2} \tau^{\dagger}T\, \tau \, ,\\
H(\theta)=\frac{1}{2\sigma_{\theta}^{2}} \int \mathrm{d}x \left( \frac{\partial^{2}\theta}{\partial^{2}x}\right)^{2}= \frac{1}{2}\theta^{\dagger}\Theta\, \theta \, .
\end{align}
The strength parameters of the curvature, $\sigma_{\theta}$ and $\sigma_{\tau}$, represent the tolerated deviation by weighting the $L_{2}$-norm of the second derivative.
For the limit $\sigma_{\theta \setminus \tau } \rightarrow \infty $ the expected roughness is very high and therefore rough realizations are not suppressed. 
Otherwise, for the limit $\sigma_{\theta \setminus \tau } \rightarrow 0 $ the smoothness prior enforces a linear behavior in the chosen coordinate systems, which correspond to a power-law for the power spectrum. 
However, the exact values of $\sigma_{\theta \setminus \tau } $ are not the driving forces for the inference.
Since the priors on $\theta$ and $\tau$ only punish curvature, there is a degeneracy in $U$.
We can add a constant factor $a \in \mathbb{R}$ to $\theta$, if we subtract at the same time  $2a$ from $\tau$. 
This operation leaves the functional form of $U$ invariant.
As one is not interested in the individual values of $\tau$ or $\theta$, this does not affect the Hamiltonian. 
One can use the degeneracy to control the numerical stability of the algorithm by keeping $\theta$ and $\tau$ in numerically stable regions.

In the following the model error parameters $(\tau, \theta)$ will be merged into the vector $q=(\tau, \theta)$. 
Thus, if we talk about parameters without further specification, both $p$ as well as $q$ are meant.
The hierarchical Bayesian network summarizing our assumptions can be seen in fig.~\ref{hierachical_model}.

Similar as for the Wiener filter theory in sec.~\ref{Wiener_filter}, we introduce the posterior information Hamiltonian $\mathcal{H}(p, \tau , \theta , u \vert d )$ and the partition function $\mathcal{Z}(d)$:
\begin{align}
\mathcal{P}(p,\tau, \theta, u|d) = \frac{1}{\mathcal{Z}(d)}e^{-\mathcal{H}(d,p,\tau, \theta, u)}, \mbox{with} \\
\mathcal{H}(d,p,\tau, \theta, u) = -\ln[\mathcal{P}(d,p,\tau, \theta, u)] \, ,\\
\mathcal{Z}(d) = \int \mathcal{D}p \,\mathcal{D}\tau \,\mathcal{D}\theta \, \mathcal{D} u \, e^{-\mathcal{H}(d,p,\tau, \theta, u)} = \mathcal{P}(d)\, . \label{partition_sum}
\end{align}
\begin{figure}[]
	\centering 
	\resizebox{\linewidth}{!}{%
	\begin{tikzpicture}[scale=0.2,every node/.style ={scale = 1.}]
	\node[place]  (s)[shape=rectangle]  {s};
	\node[place]  (n)[shape=rectangle,right=2cm of s] {$n$};
	\node[place]  (d=Rs+n)[shape=rectangle,inner sep = 2pt, below=2cm of n] {$d=R\,s+n$};
	\node[place]  (Gn)[shape=rectangle, inner sep = 2pt , above=2cm of n] {$\mathcal{G}(n,N)$};
	\node[place]  (n1)[shape=rectangle, above left =0.5cm and 0.3 of n] {$n_{1}$};
	\node[place]  (n2)[shape=rectangle, above right =0.5cm and 0.3 of n] {$n_{m}$};		  
	\node[place] (n5)[shape=rectangle, draw=white, right=0.03cm of n1]{.....};
	\node[place] (n6)[shape=rectangle, draw=white, left=0.03cm of n2]{.....};
	\node[place]  (n3)[shape= coordinate, below= 0.002cm of n1]{};
	\node[place]  (n4)[shape= coordinate, below= 0.002cm of n2]{};
	\draw[thick,decorate,decoration={brace,mirror,amplitude=12pt}] (n3) -- (n4) node[midway, above,yshift=-14pt,]{};
	\node[place]  (u)[shape=rectangle, above=1.5cm of Gn] {$u$};
	\node[place]  (Gu)[shape=rectangle, inner sep = 2pt, above=2cm of u] {$\mathcal{G}(u,U(q))$};
	\node[place]  (u1)[shape=rectangle, above left =0.5cm and 0.3 of u] {$u_{1}$};
	\node[place]  (u2)[shape=rectangle, above right =0.5cm and 0.3 of u] {$u_{m}$};		  
	\node[place] (u5)[shape=rectangle, draw=white, right=0.03cm of u1]{.....};
	\node[place] (u6)[shape=rectangle, draw=white, left=0.03cm of u2]{.....};
	\node[place]  (u3)[shape= coordinate, below= 0.002cm of u1]{};
	\node[place]  (u4)[shape= coordinate, below= 0.002cm of u2]{};
	\draw[thick,decorate,decoration={brace,mirror,amplitude=12pt}] (u3) -- (u4) node[midway, above,yshift=-14pt,]{};
	\node[place]  (q)[shape=rectangle, above=1cm of Gu]{$q$};
	\node[place]  (tau)[shape=rectangle, above left=0.5cm and 1cm of q]{$\tau$};
	\node[place]  (theta) [shape=rectangle, above right=0.5cm and 1cm of q]{$\theta$};
	\node[place]  (tau1)[shape= coordinate, below= 0.002cm of tau]{};
	\node[place]  (theta1)[shape= coordinate, below= 0.002cm of theta]{};
	\draw[thick,decorate,decoration={brace,mirror,amplitude=12pt}] (tau1) -- (theta1) node[midway, above,yshift=-14pt,]{};
	\node[place]  (Gtau) [shape=rectangle, inner sep = 2pt, above=1cm of tau]{$\mathcal{G}(\tau,T)$}; 
	\node[place]  (Gtheta)[shape=rectangle, inner sep = 2pt, above=1cm of theta]{$\mathcal{G}(\theta,\Theta)$};
	\node[place]  (tp)[shape=rectangle, above left=2cm and 2cm of s]{$t(p)$};
	\node[place]  (p)[shape=rectangle, above=1.5cm of tp]{$p$};
	\node[place]  (Pp)[shape=rectangle, above=2cm of p] {$\mathcal{P}(p)$};
	\node[place]  (p1)[shape=rectangle, below left =0.7cm and 0.3 of Pp] {$p_{1}$};
	\node[place]  (p2)[shape=rectangle, below right =0.7cm and 0.3 of Pp] {$p_{m}$};		  
	\node[place] (p5)[shape=rectangle, draw=white, right=0.01cm of p1]{......};
	\node[place] (p6)[shape=rectangle, draw=white, left=0.01cm of p2]{......};
	\node[place]  (p3)[shape= coordinate, below= 0.002cm of p1]{};
	\node[place]  (p4)[shape= coordinate, below= 0.002cm of p2]{};
	\node[place]  (sigmatau)[shape=rectangle, above left =0.5cm and 0.5 cm of Gtau] {$\sigma_{\tau}$};  
	\node[place]  (sigmatheta)[shape=rectangle, above right =0.5cm and 0.5cm of Gtheta] {$\sigma_{\theta}$};
	\draw[thick,decorate,decoration={brace,mirror,amplitude=12pt}] (p3) -- (p4) node[midway, above,yshift=-14pt,]{};
	\draw[->] (Gn) to (n5);
	\draw[->] (Gn) to (n6);
	\draw[->] (s) to (d=Rs+n) ;
	\draw[->] (n) to (d=Rs+n);
	\draw[->] (u) to (s);
	\draw[->] (Gu) to (u5);
	\draw[->] (Gu) to (u6);
	\draw[->] (Gu) to (u1);
	\draw[->] (Gu) to (u2);
	\draw[->] (Gtau) to (tau);
	\draw[->] (Gtheta) to (theta);
	\draw[->] (p) to (tp);
	\draw[->] (tp) to (s);
	\draw[->] (q) to (Gu);
	\draw[->] (Gn) to (n1);
	\draw[->] (Gn) to (n2);
	\draw[->] (Pp) to (p5);
	\draw[->] (Pp) to (p6);
	\draw[->] (Pp) to (p1);
	\draw[->] (Pp) to (p2);
	\draw[->]	(sigmatheta) to (Gtheta);
	\draw[->]	(sigmatau) to (Gtau);
	\end{tikzpicture}
}
	\caption{Graphical model of the hierarchical Bayesian network introduced in sec.~\ref{model_error_model}. The parameters in the top row, $\sigma_{\tau}$ and $ \sigma_{\theta} $, as well as $\mathcal{P}(p)$ have to be specified by the user. The parameters characterizing the model error uncertainty $\theta, \tau$ and the model parameters $p$ will be infered by the algorithm.}\label{hierachical_model}
\end{figure}
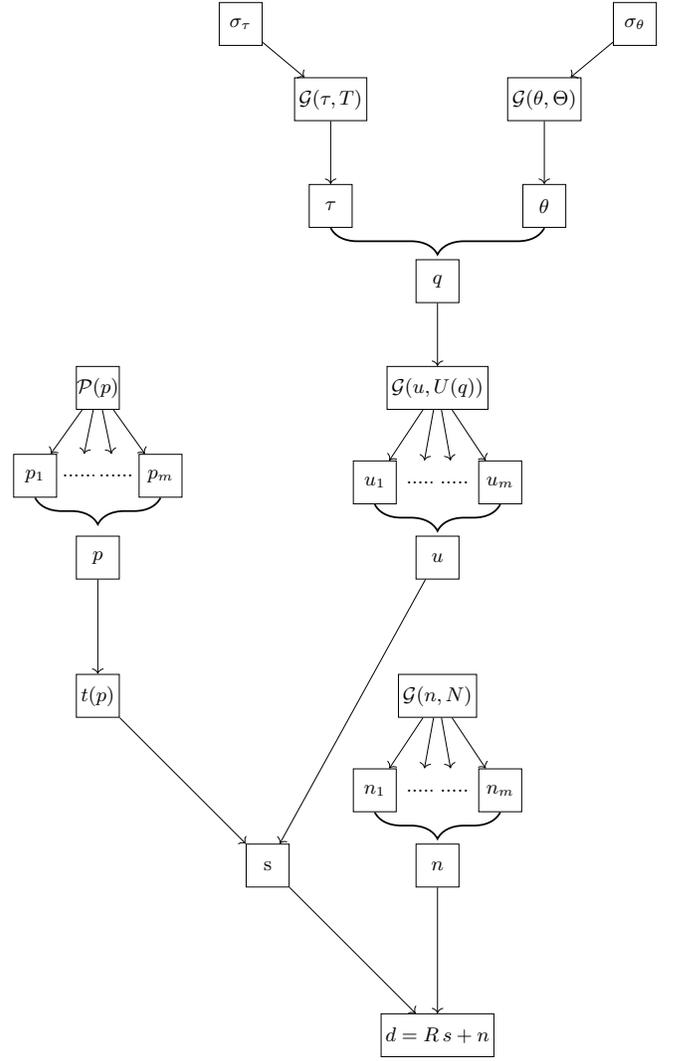
\begin{figure}[]

	\tikzset{every picture/.style={line width=0.75pt}} 
	
	\begin{tikzpicture}[x=0.75pt,y=0.75pt,yscale=-0.45,xscale=0.45]
	
	\draw    (140, 260) circle [x radius= 25, y radius= 25]  ;
	\draw    (88, 164) circle [x radius= 25, y radius= 25]  ;
	\draw    (192, 164) circle [x radius= 25, y radius= 25]  ;
	\draw    (133, 57) circle [x radius= 25, y radius= 25]  ;
	\draw    (243, 57) circle [x radius= 25, y radius= 25]  ;
	\draw    (265,166) -- (362,166.98) ;
	\draw [shift={(364,167)}, rotate = 180.58] [color={rgb, 255:red, 0; green, 0; blue, 0 }  ][line width=0.75]    (10.93,-3.29) .. controls (6.95,-1.4) and (3.31,-0.3) .. (0,0) .. controls (3.31,0.3) and (6.95,1.4) .. (10.93,3.29)   ;
	
	\draw    (534, 232) circle [x radius= 25, y radius= 25]  ;
	\draw    (435, 123) circle [x radius= 25, y radius= 25]  ;
	\draw    (494, 93) circle [x radius= 25, y radius= 25]  ;
	\draw    (575, 93) circle [x radius= 25, y radius= 25]  ;
	\draw    (97,187) -- (125,235.27) ;
	\draw [shift={(126,237)}, rotate = 239.89] [color={rgb, 255:red, 0; green, 0; blue, 0 }  ][line width=0.75]    (10.93,-3.29) .. controls (6.95,-1.4) and (3.31,-0.3) .. (0,0) .. controls (3.31,0.3) and (6.95,1.4) .. (10.93,3.29)   ;
	
	\draw    (180,186) -- (152.99,233.26) ;
	\draw [shift={(152,235)}, rotate = 299.74] [color={rgb, 255:red, 0; green, 0; blue, 0 }  ][line width=0.75]    (10.93,-3.29) .. controls (6.95,-1.4) and (3.31,-0.3) .. (0,0) .. controls (3.31,0.3) and (6.95,1.4) .. (10.93,3.29)   ;
	
	\draw    (143,79) -- (180.91,137.32) ;
	\draw [shift={(182,139)}, rotate = 236.98] [color={rgb, 255:red, 0; green, 0; blue, 0 }  ][line width=0.75]    (10.93,-3.29) .. controls (6.95,-1.4) and (3.31,-0.3) .. (0,0) .. controls (3.31,0.3) and (6.95,1.4) .. (10.93,3.29)   ;
	
	\draw    (232,79) -- (203.88,136.21) ;
	\draw [shift={(203,138)}, rotate = 296.18] [color={rgb, 255:red, 0; green, 0; blue, 0 }  ][line width=0.75]    (10.93,-3.29) .. controls (6.95,-1.4) and (3.31,-0.3) .. (0,0) .. controls (3.31,0.3) and (6.95,1.4) .. (10.93,3.29)   ;
	
	\draw    (454,139) -- (510.79,213.41) ;
	\draw [shift={(512,215)}, rotate = 232.65] [color={rgb, 255:red, 0; green, 0; blue, 0 }  ][line width=0.75]    (10.93,-3.29) .. controls (6.95,-1.4) and (3.31,-0.3) .. (0,0) .. controls (3.31,0.3) and (6.95,1.4) .. (10.93,3.29)   ;
	
	\draw    (501,116) -- (524.48,203.07) ;
	\draw [shift={(525,205)}, rotate = 254.91] [color={rgb, 255:red, 0; green, 0; blue, 0 }  ][line width=0.75]    (10.93,-3.29) .. controls (6.95,-1.4) and (3.31,-0.3) .. (0,0) .. controls (3.31,0.3) and (6.95,1.4) .. (10.93,3.29)   ;
	
	\draw    (569,116) -- (542.58,203.09) ;
	\draw [shift={(542,205)}, rotate = 286.88] [color={rgb, 255:red, 0; green, 0; blue, 0 }  ][line width=0.75]    (10.93,-3.29) .. controls (6.95,-1.4) and (3.31,-0.3) .. (0,0) .. controls (3.31,0.3) and (6.95,1.4) .. (10.93,3.29)   ;
	
	\draw    (617,143) -- (557.31,211.49) ;
	\draw [shift={(556,213)}, rotate = 311.07] [color={rgb, 255:red, 0; green, 0; blue, 0 }  ][line width=0.75]    (10.93,-3.29) .. controls (6.95,-1.4) and (3.31,-0.3) .. (0,0) .. controls (3.31,0.3) and (6.95,1.4) .. (10.93,3.29)   ;
	
	\draw    (634, 125) circle [x radius= 25, y radius= 25]  ;
	
	\draw (133,57) node  [align=left] {};
	\draw (434,122) node [scale=1.44]  {$\xi $};
	\draw (140,259) node [scale=1.44]  {$d$};
	\draw (88,162) node [scale=1.44]  {$p$};
	\draw (191,162) node [scale=1.44]  {$s$};
	\draw (131,55) node [scale=1.44]  {$\theta $};
	\draw (242,56) node [scale=1.44]  {$\tau $};
	\draw (492,94) node [scale=1.44]  {$\theta $};
	\draw (575,93) node [scale=1.44]  {$\tau $};
	\draw (636,125) node [scale=1.44]  {$p$};
	\draw (533,234) node [scale=1.44]  {$d$};	
	\end{tikzpicture}
	
	\caption{Causality structure of the classical formulation of our problem
		(left) and of its reformulation (right)}\label{systematics_variable_change}
\end{figure}
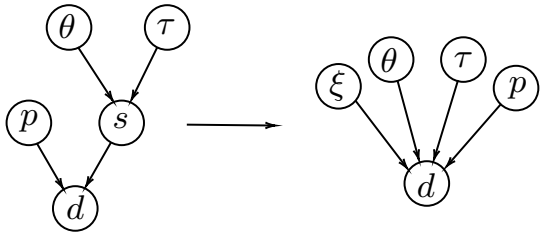
The resulting full Hamiltonian can be calculated as before, such that we get
\begin{equation}
\begin{aligned}
\mathcal{H}(p,\tau, \theta , u , d) &= \frac{1}{2} \left\{ d-e^{R[t(p)+u]}  \right\}^{\dagger}N^{-1}\left\{d-e^{R[t(p)+u]}\right\} \\ 
&+\frac{1}{2} \ln(\vert 2 \pi N \vert) + H(p) +\frac{1}{2}u^{\dagger}U^{-1}u  \\
&+\frac{1}{2}\ln(\vert 2 \pi U \vert) +\frac{1}{2}\tau^{\dagger}T\, \tau  +\frac{1}{2}\theta^{\dagger}\Theta\,\theta \\
&+\frac{1}{2}\ln(\vert 2 \pi T \vert)+\frac{1}{2}\ln(\vert 2 \pi \Theta \vert) \, .
\end{aligned}
\end{equation}
The model error $u$ is assumed to be generated by drawing from a Gaussian distribution, which is neither diagonal in position nor in Fourier space. 
Furthermore, it is possible to draw from a Gaussian distribution in its eigenbasis by weighting an independent, white spectrum excitation field $\xi$ with unit variance with the square root of its eigenvalues. 
We will adopt this idea and reformulate our problem in the following way.
\begin{eqnarray}
\begin{aligned}
u &= A\xi \\
A&=\widehat {e^{{\theta_{x}}}}\mathbb{F}^{\dagger}\widehat{ \mathbb{P}^{\dagger} e^{{\frac{1}{2}\tau_{y}}}} \\
\mathrm{s.t.} \;\;\; A A^{\dagger} &= U \, .
\end{aligned}
\end{eqnarray}
Here we introduced the amplitude operator $A$. 
The change in causality can be seen in fig.~\ref{systematics_variable_change}, where one observes that now all unknown fields directly interact with the data.
This reparametrization is following the ideas of~\cite{Repara,Jakobneu}.
Inserting the change of variables from $u$ to $\xi$ into the Hamiltonian gives
\begin{equation}
\begin{aligned}
\mathcal{H}(p,\tau, \theta , \xi ,  d ) &= \frac{1}{2}\left\{d-Re^{[t(p)+A\, \xi]}\right\}^{\dagger}N^{-1}  \\
&\times\left\{d-Re^{[t(p)+A\, \xi]}\right\} + \mathcal{H}(p) \\
 &+\frac{1}{2}\xi^{\dagger} \mathbbm{1}\, \xi+\frac{1}{2}\theta^{\dagger}\Theta\,\theta + \frac{1}{2}\tau^{\dagger}T\,\tau +\mathcal{H}_{0} \, .
\end{aligned}
\end{equation}
All terms that are unimportant for the later process are collected in $\mathcal{H}_{0}$:
\begin{equation}
\mathcal{H}_{0}= \frac{1}{2}\ln(\vert 2 \pi T \vert)+\frac{1}{2}\ln(\vert 2 \pi \Theta \vert) + \frac{1}{2} \ln(\vert 2 \pi N \vert) +\frac{1}{2}\ln(\vert 2\pi \mathbbm{1}\vert)\, . 
\end{equation}
These terms will not affect the further inference.
For the inference we will need the posterior Hamiltonian
\begin{equation}
\mathcal{H}(p,\tau, \theta , \xi \vert  d )=  \mathcal{H}(p,\tau, \theta , \xi , d ) + \ln \left[ Z(d) \right] \, ,
\end{equation}
with $\ln \left[ Z(d) \right]$ defined in eq.~\ref{partition_sum}.
Since the partition sum does not depend on $\xi$, $q$ and $p$, we can ignore it in the following.
\section{Method}\label{method}
In this section we want to present the approximation of the posterior distribution via variational Bayes. 
%
%
Thus, we adapt the methods presented in~\cite{Global-Newton,2017arXiv171102955K} which performs a minimization of all unknown fields at the same time and provides a posterior uncertainty estimation via samples.
\subsection{Approximation of the posterior distribution}
To use the implementation in the Numerical Information field theory package (NIFTy) \cite{Nifty,Nifty3}, we have to ensure that all used priors, except the noise prior, are white priors.
Due to this fact, we have to reformulate our parameterization of $U$ in eq.~\ref{U_para}.
As already mentioned the reformulation is possible by weighting a white excitation field with the square root of the covariance. 
This can be done for the smoothness priors on $\theta$ and $\tau$. 
\begin{equation}
\begin{aligned}
\theta &\rightarrow A_{\theta} \xi_{\theta} \\
\tau &\rightarrow A_{\tau}\xi_{\tau} \\
A_{\theta} = \reallywidehat{e^{\theta}} &\rightarrow  \reallywidehat{e^{\widetilde{{A_{\theta}}}\xi_{\theta}}}\\
A_{\tau} = \reallywidehat{ \mathbb{P}^{\dagger}e^{\tau}} &\rightarrow \reallywidehat{\mathbb{P}^{\dagger}e^{\widetilde{{A_{\tau}}}\xi_{\tau}}}
\end{aligned}
\end{equation}
The exact transformations and details on the chosen coordinates for the excitation fields are given in appendix~\ref{app: trafos}. 
Thus, the covariance operator $U$ has the reformulated form
\begin{equation}
U= A_{\theta} \mathbb{F}^{\dagger} A_{\tau}^{\dagger} A_{\tau} \mathbb{F} A_{\theta} \, .
\end{equation}
Now, we can again use a variable transformation to write 
\begin{equation}
\begin{aligned}
u = A\xi \, \, ,\mathrm{with} \\
A = A_{\theta}\mathbb{F}^{\dagger}A_{\tau} \, .
\end{aligned}
\end{equation}
The coordinate transformation of the parameters $p$ into a coordinate system $\xi_{p}$ where the prior is white is a little bit more subtle, since the prior for the parameters can almost have any form.
However, as long as the cumulative distribution function of the prior can be calculated analytically, this coordinate transformation can always be applied~\cite{Repara,Jakobneu}. \\
As a next step one stores all unknown fields, namely $\xi, \xi_{\theta}, \xi_{\tau}, \xi_{p}$, in a multi-field $\phi$.
Now, we can continue as before and calculate the full Hamiltonian
\begin{equation}
\begin{aligned}
\mathcal{H}(\xi_{p},\xi_{\tau}, \xi_{\theta} , \xi_{\phi} ,\xi ,  d ) &= \sum_{i = 1}^{N_{\mathrm{data}}} \left[  \frac{1}{2}\left\{d_{i}-Re^{[t(\xi_{p_{i}})+A\, \xi_{i}]}\right\}^{\dagger}N^{-1} \right.\\
& \times\left\{d_{i}-Re^{[t(\xi_{p_{i}})+A\, \xi_{i}]}\right\}  +\frac{1}{2}\xi_{i}^{\dagger} \mathbbm{1}\, \xi_{i}  \\
&+ \left. \frac{1}{2}\xi_{p_{i}}^{\dagger}\mathbbm{1}\xi_{p_{i}} \right] 
+\frac{1}{2}\xi_{\theta}^{\dagger}\mathbbm{1}\xi_{\theta} \\
&+ \frac{1}{2}\xi_{\tau}^{\dagger}\mathbbm{1}\,\xi_{\tau} +\mathcal{H}_{0} \, .
\end{aligned}
\end{equation}
Usually one would need to calculate the derivatives of this in order to be able to apply a Newton scheme. 
However, since the model for the model error is constructed from NIFTy model objects, NIFTy is able to compute all derivatives regarding $\xi, \xi_{\theta}, \xi_{\tau}$ on its own.
Only the gradient with respect to the model parameters $\xi_{p}$ has to be calculated and implemented, which is individual for each model $t(\xi_{p})$.
\\
In order to apply a variational inference method, the posterior distribution has to be approximated. 
The advantage of the method implemented in NIFTy is that we can approximate the posterior as a Gaussian in all unknown fields
\begin{equation}
\widetilde{\mathcal{P}}(\phi \vert d ) = \mathcal{G}(\phi - \bar{\phi}, \Phi) \, ,
\end{equation}
where $\bar{\phi}$ represents the mean of the multi-field $\phi$ and $\Phi$ is its full covariance matrix, which contains the crosscorrelations of the different fields within the multi-field. 
By using this approximation one can minimize the variational Kullback-Leibler divergence ($D_{\mathrm{KL}}$), which is given by 
\begin{equation}
\begin{aligned}
D_{\mathrm{KL}} &= \int \mathcal{D}\phi \widetilde{\mathcal{P}}(\phi \vert d) \ln \left[ \frac{\widetilde{\mathcal{P}}(\phi \vert d)}{\mathcal{P}(\phi \vert d)}\right] \\ &= \langle \mathcal{H}(\phi \vert d) \rangle_{\mathcal{G}(\phi - \bar{\phi} , \Phi)} - \frac{1}{2} \ln \left( \vert 2 \pi e \Phi \vert \right) \, .
\end{aligned}
\end{equation}
\subsection{Numerical example}
\begin{figure}
	\centering
	\includegraphics[width=\columnwidth, height = 0.35\textheight]{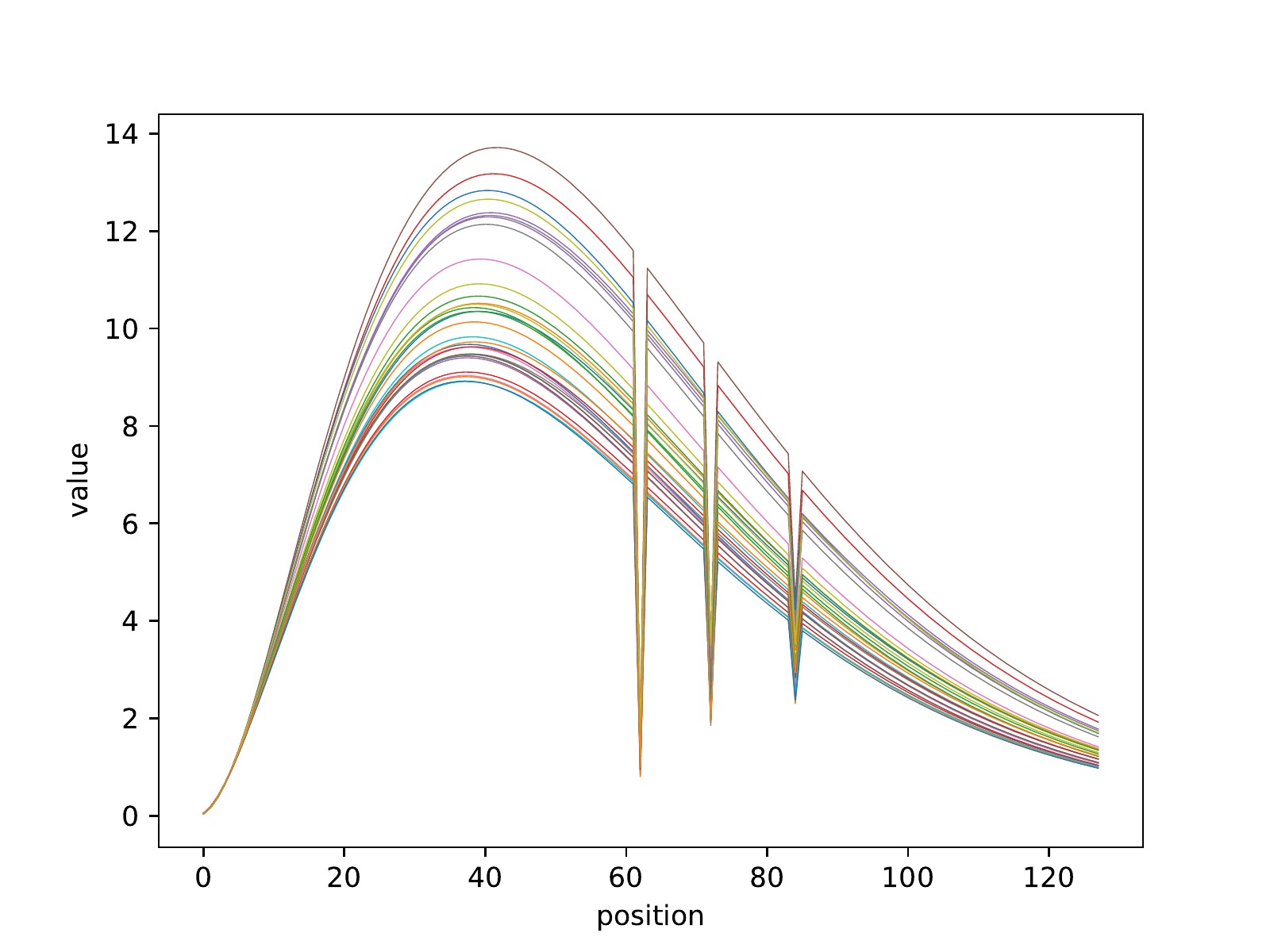}
	\caption{32 different models with individual parameter values, which are used in the inference for the model errors and parameters.}\label{different_models}
\end{figure}

\begin{figure}[]
	\centering
	\includegraphics[width=\columnwidth, height = 0.35\textheight]{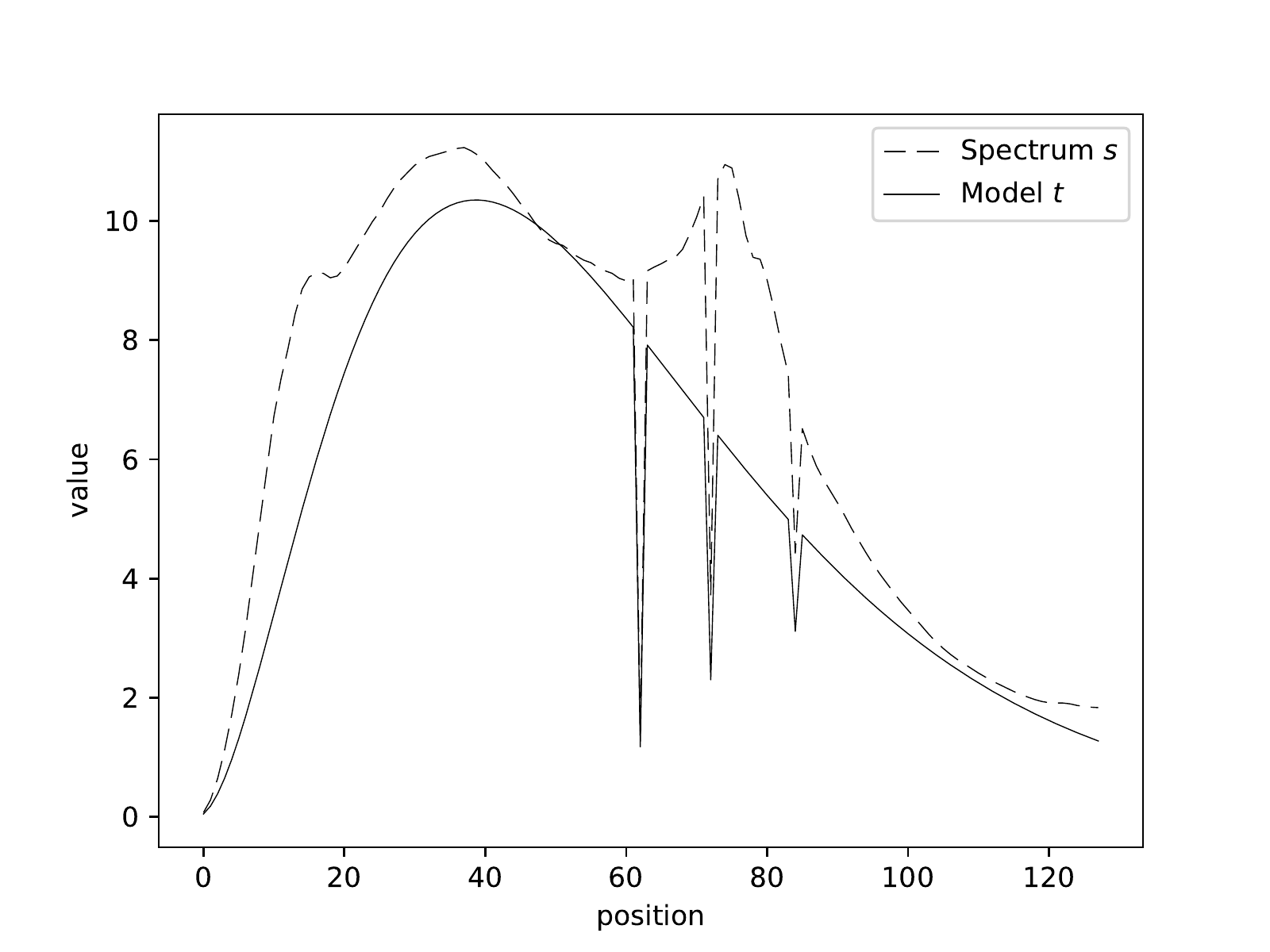}
	\caption{One of the spectra with the corresponding model.}\label{error-signal}
\end{figure}

\begin{figure}
	\centering
	\includegraphics[width=\columnwidth, height = 0.35\textheight]{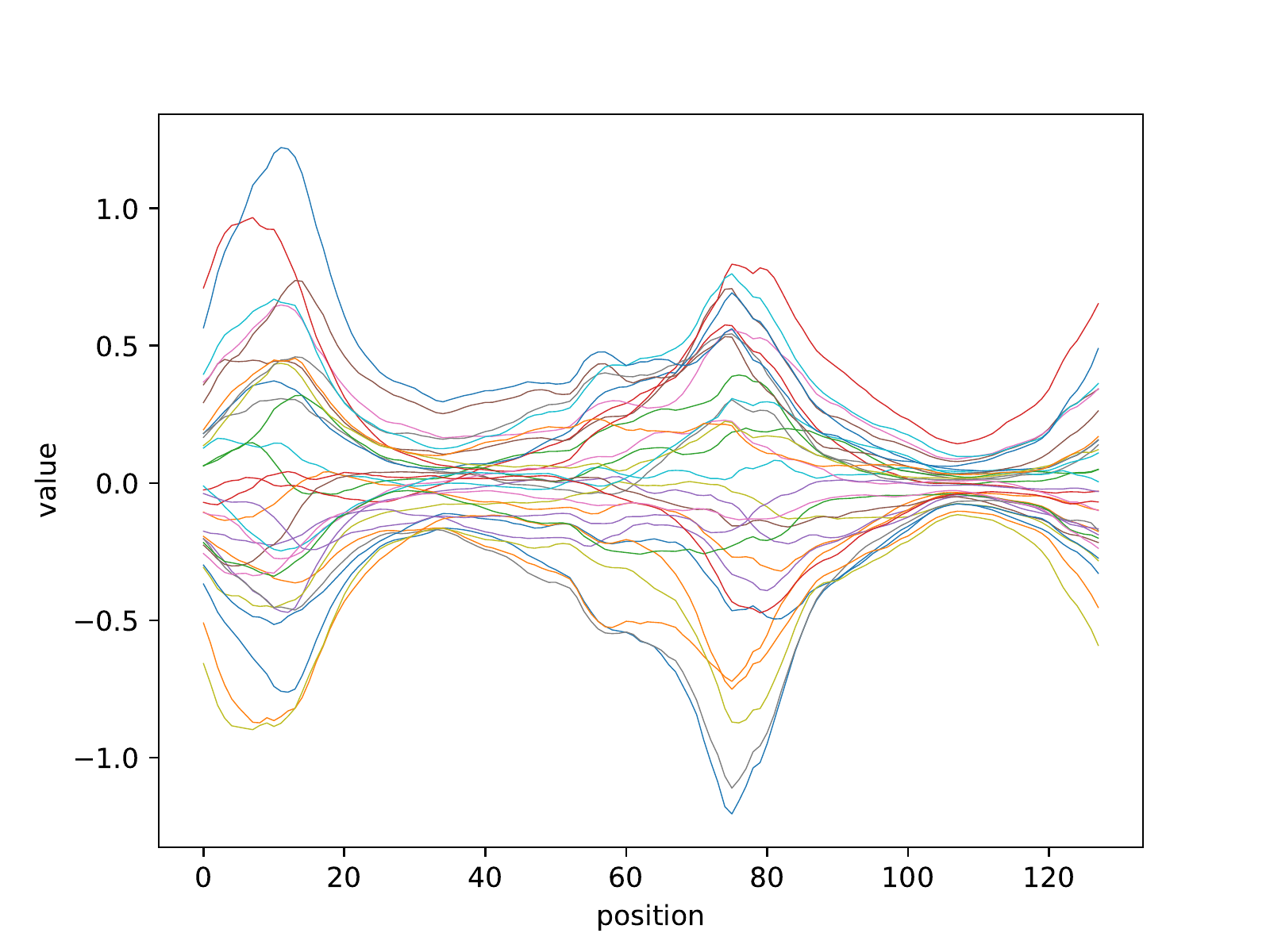}
	\caption{32 different model error realizations drawn from $U$}\label{different_errors}
\end{figure}

\begin{figure}[]
	\centering
	\includegraphics[width=\columnwidth, height = 0.35\textheight]{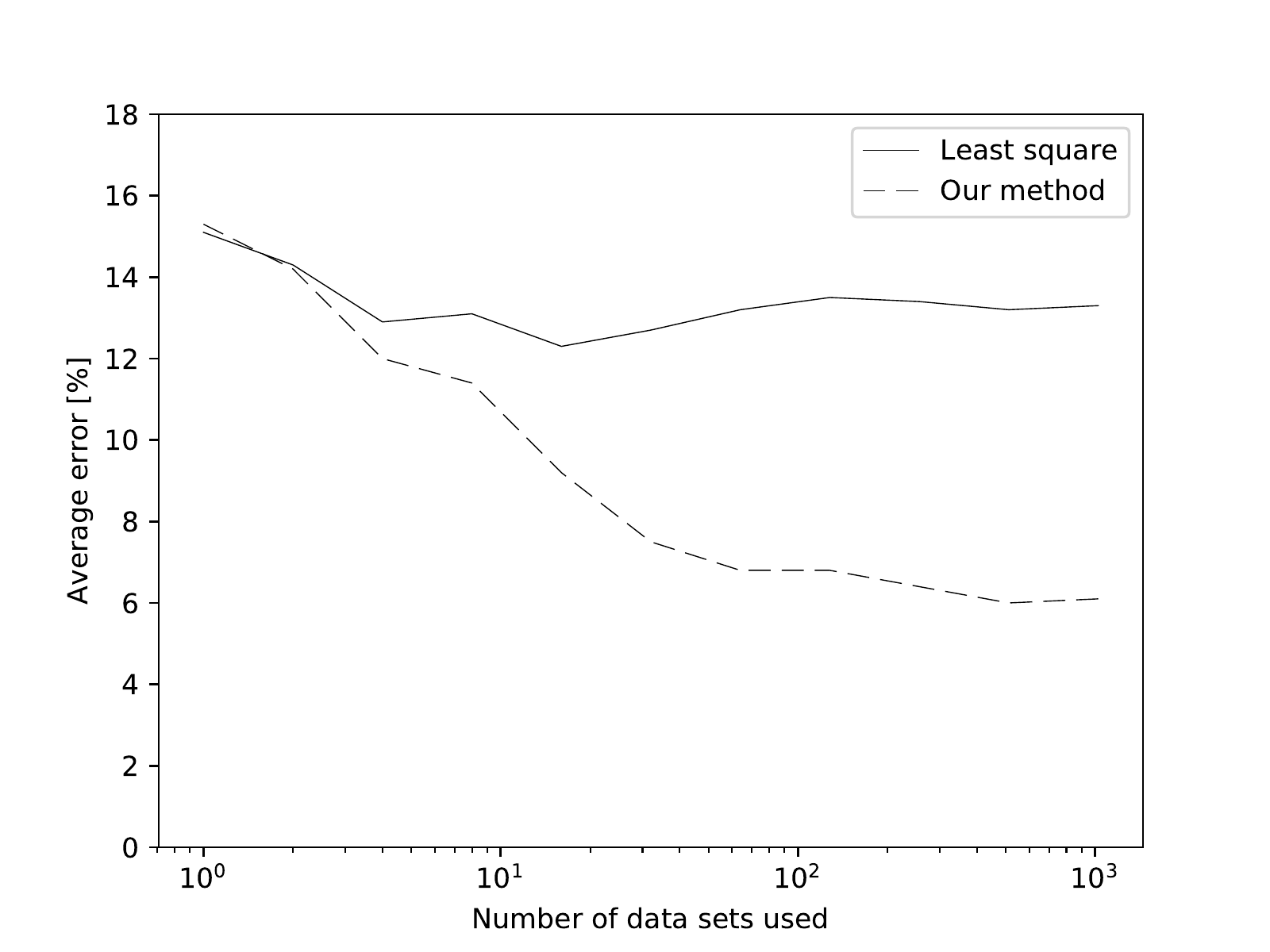}
	\caption{Comparison of the errors of a least-square parameter estimation and our method for different number of data sets.}\label{compare}
\end{figure}

\begin{figure}[]
	\centering
	\includegraphics[width=\columnwidth, height = 0.35\textheight]{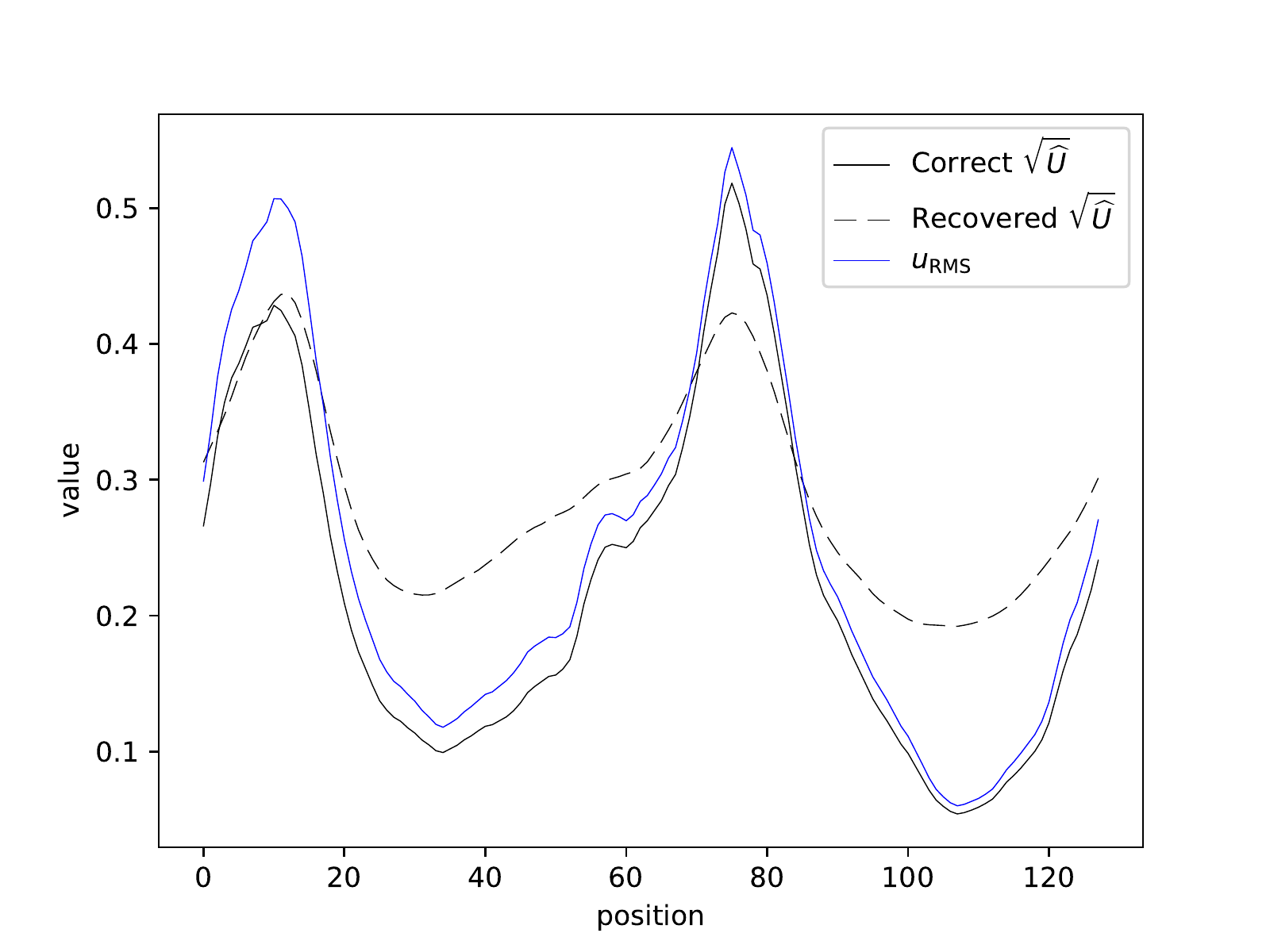}
	\caption{The correct square root of the diagonal of U and its reconstruction together with the positive average error are shown.}\label{u_U_combination}
\end{figure}

We demonstrate the applicability of our derived algorithm for one-dimensional synthetic data. 
The algorithm is implemented in python using the NIFTy package. 
By comparing our algorithm with the least-square algorithm, we show its capability. 
In our example the signal has a resolution of 128 pixels.  
As a prior for the parameters we used a Gaussian, such that
\begin{equation}
\mathcal{H}(p_{i}) = \frac{1}{2}(p_{i}-p^{\ast})P^{-1}(p_{i}-p^{\ast}) \, ,
\end{equation}
where we assumed the prior matrix $P^{-1}$ to be diagonal.
As discussed before, we have to transform the parameters, such that they have a white prior.
For the chosen prior this is straight forward and the transformation is given by
\begin{equation}
p_{i} \rightarrow p^{\ast}_{i} + \sigma_{i}\xi_{p_{i}}\, ,
\end{equation}
where $p^{\ast}_{i}$ is the mean estimation of the $i$-th parameter and $\sigma_{i} = \widehat{P}^{\frac{1}{2}}$ represents the expected deviation of the mean. 
Here the $\widehat{\; \; \; \; \;}$ denotes the diagonal of an operator.
We have chosen $\sigma_{i} =\left( 0.3, 1., 0.8, 0.5,0.3\right)$ and the mean of each data set was drawn uniformly from ranges the $ \left([1.9,2.1], [19.6,20.4],[1.8,2.2],[0.8,1.2],[0.4,0.6] \right)$.
Thus, we can draw a excitation field $\xi_{p}$ from the white prior.
The corresponding models which are used in the example with 32 different data sets and models can be seen in fig. \ref{different_models}.\\
We produce the data sets according to eq.~\ref{real_data_equation}, where the signal field $s$ is a sum of the error and the model error.  \\
By drawing a white excitation field $\xi_{\theta}$ and fixing the power-spectrum to 
\begin{equation}
P_{\xi_{\tau}}(k)= \frac{1}{(k^{2}+1)}\, ,
\end{equation}
we construct the operator $A$. 
The noise covariance is given by $0.2 \times \mathbbm{1}$ and the measurement instrument is a unit response.
The gradient of the model with respect to the transformed parameters $\xi_{p_{i}} = \left(\xi_{\alpha_{i}}, \xi_{T_{\mathrm{eff}_{i}}}, \xi_{t_{1_{i}}}, \xi_{t_{2_{i}}}, \xi_{t_{3_{i}}}\right)$ reads 
\begin{equation}\label{eq: derivatives_model3}
t'(\xi_{p_{i}}) = \begin{bmatrix}
\sigma_{\alpha}\ln( \frac{\nu}{\nu_{0}}) \\
\frac{\sigma_{T_{\mathrm{eff}}}}{\left(\mu_{T_{\mathrm{eff}_{i}}} + \sigma_{T_{\mathrm{eff}}}\xi_{T_{\mathrm{eff}_{i}}}\right)^{2}} \\
-\sigma_{t_{1}}\mathcal{G}(\nu-\nu_{1},\sigma_{1}^{2})\\
-\sigma_{t_{2}}\mathcal{G}(\nu-\nu_{2},\sigma_{2}^{2}) \\
-\sigma_{t_{3}}\mathcal{G}(\nu-\nu_{3},\sigma_{3}^{2})
\end{bmatrix} .
\end{equation}
Fig.~\ref{error-signal} shows one of the spectra with the corresponding model. 
The model with its free parameters is not able to capture all the aspects of the spectrum.
Conventional algorithms would try to explain the differences through the model, whether they are due to the model error or to the model.
In fig.~\ref{different_errors} the 32 different model errors and the square root of the covariance $U$ are shown, which illustrates the correlation structure and the difficulty to reconstruct a spatial varying uncertainty covariance. 
One can observe that for this special $U$ the model error mainly imprints on the spectra in their ascending parts and the part where the absorption lines are present.\\
Fig. \ref{compare} shows a comparison on between our method and a least-square method (LS) for the estimation of model parameters. 
Using multiple records can cause the optimization scheme to go to a local minimum.
After some tests, we decided to divide the likelihood Hamiltonian by the number of data sets used in the particular inference similar to the proposed solution of this problem by~\cite{boudjemaa2004parameter}..
We can observe that our method is able to learn from an increasing number of data sets while the LS estimates each data set on its own and thus does not improve with more data sets. 
However, our method saturates at about 512 data sets on an average parameter error of $6\%$ and the addition of more data sets does not further reduce the error.
As for each model still the parameters $p_{i}$ have to be found individually in the presence of the unknown noise relations $n_{i}$ and $u_{i}$. \\
The square root of the diagonal of the model error uncertainty covariance $\sqrt{\widehat{U}}$ and its reconstruction together with the root mean square error $u_{\mathrm{RMS}}$ are shown in fig.~\ref{u_U_combination}.
One would expect that the recovered uncertainty, as an conservative estimator, is in general smaller than the correct uncertainty.
However, in our reconstruction of the model error uncertainty the peaks can be modeled well, while the tails are overestimated. \\
\section{Conclusion}
Estimations of parameters from models are essential to get insights into physics. 
However, many physical objects are not understood in their full complexity. 
Thus, the model errors need to be included in the error uncertainty, but this would afford a case by case analysis of the missing physics and therefore is often impractical.
To address these problems this paper constructs a plausible and effective description of model errors and their uncertainty.
To constraint the additional degrees of freedom it relies on using several data sets at the same time in a joint inference. \\
We derive an algorithm to reconstruct the parameters of a model as well as the model error and its statistics from noisy data.
By using more than one data set of the same type of stars, we got a more sophisticated estimate for the model error uncertainty.
It was important to build a very general approximating description for the model error uncertainty in order to be capable for most sorts of model errors. 
One can see that a few general assumptions about the nature of model errors, i.e. locality and smoothness, can be sufficient to reconstruct the model error uncertainty when one has access to a large set of data.
The algorithm can be graphically visualized through a Bayesian hierarchical model.
It was able to reconstruct the model errors as well as the model parameters.\\
Additionally, expectation means can be calculated through the samples from the inverse metric. 
By using approximate posterior samples the opportunity to deal with more complex uncertainty structures is computationally feasible.
We can conclude that this paper enables the opportunity to deal with parameter estimates of miss-specified models by accounting for the model error and its properties within a Bayesian approach.
\bibliography{bibliography} 
\bibliographystyle{ieeetr}
\appendix
\section{Definitions}
\subsection{Symbolic conventions}\label{definitions}
The scalar product between a field $a^{\dagger}$ and another field $b$ is defined in the following way
\begin{equation}
a^{\dagger}b = \int \mathrm{d} x \, a^{\ast}(x) b(x) \, .
\end{equation}
An operator $A$ acting on a field $b$ is defined as 
\begin{equation}
\left(A\,b\right)(x') = \int \mathrm{d} x \, A(x',x)\, b(x)\, .
\end{equation}
\subsection{Power projector}\label{power_proj}
The power projection operator averages all positions in the Fourier space  belonging to one spectral bin.
Let $p_{k}$ be a component of a field over the power space.
Then the corresponding component $f_{k}$ of the field over the Fourier space can be calculated by
\begin{equation}\label{eq : Power_projector}
p_{k}= \mathbb{P}_{k\kappa}f_{\kappa}=\frac{1}{\rho_{k}}\int_{\kappa \in k} \frac{\mathrm{d}\kappa}{(2\pi)^{\mathrm{dim}}} f_{\kappa} \, ,
\end{equation}
where $\rho_{k}$ is the bin volume and $\mathrm{dim}$ indicates the dimensionality of $f$. \\
\subsection{Transformation to a white prior}\label{app: trafos}
Since a smoothness prior is best represented in Fourier space, where it is just a multiplication with the factor $\frac{1}{k^{4}}$, $\theta$ will be transformed to a field $\xi_{\theta}$, which lives in Fourier space.
Thus, the transformation of $\theta$ to $\xi_{\theta}$ is given by
\begin{equation}
\theta \rightarrow \mathbb{F}^{\dagger}\, \widehat{\mathbb{P}^{\dagger}_{\mathrm{lin}} \, P_{\xi_{\theta}}(k)} \xi_{\theta}\, ,
\end{equation}
where $\mathbb{P}_{\mathrm{lin}}$ is a power projector as defined in eq.~\ref{eq : Power_projector} and $P_{\xi_{\theta}}(k)=\frac{\sigma_{\theta}}{k^{2}+1}$ indicates the power spectrum representing the spectral smoothness. 
The subscript $_{\mathrm{lin}}$ indicates that the power space has linear bin bounds.
We will denote the exponential of $\theta$ as
\begin{equation}
A_{\theta} = \reallywidehat{\exp\left[\theta_{x}\right]} = \reallywidehat{\exp \left[ \mathbb{F}^{\dagger}\, \widehat{\mathbb{P}^{\dagger}_{\mathrm{lin}} \, P_{\xi_{\theta}}(k)} \xi_{\theta}\right]}\, .
\end{equation}
A reformulation for the convolutional part of the model error uncertainty $U$ already exists in NIFTy.
This reformulation splits the $\tau$ field into two different fields, one describing the slope $\xi_{\phi}$ and the other one describing the deviation of a linear slope $\xi_{\tau}$. 
This model works with white priors and thus is used here.
\end{document}